\documentstyle[12pt,epsf]{article}
\input epsf
\begin{document}

\title{CONTRIBUTION OF THE NN INELASTICITY INTO
PHOTODISINTEGRATION OF THE DEUTERON}
\date{}
\author {   V.A. SADOVNIKOVA  \\
  {\it PNPI, Gatchina, 188350 St.Petersburg}}
\maketitle

\begin{abstract}
The technique of dispersion integration over the mass of
composite particle is used to describe the reaction of the deuteron
photodisintegration.  The influence of the final state interaction
(FSI) on the total cross section, calculated with and without
inelasticity is investigated.  Numerical results depend on the choice
of the vertex function for the isobar photoproduction.
\end{abstract}

The reaction of photodisintegration of the deuteron
has been considered in the framework of the dispersion relation
approach developed by V.V.Anisovich et al.$^1$. This method allows us
to construct  relativistic and gauge invariant amplitude of the
photodisintegration. Besides, in the dispersion technique the
condition $p_i^2=m_i^2$ is fulfilled for all  particles
in the intermediate state, thus no problem with the
determination of the off-mass-shell amplitudes.

To describe photodisintegration amplitude the  diagrams
shown in Fig.1 have been used.

\begin{figure}[h]
\begin{center}
\mbox{\epsfxsize=12cm \epsfbox{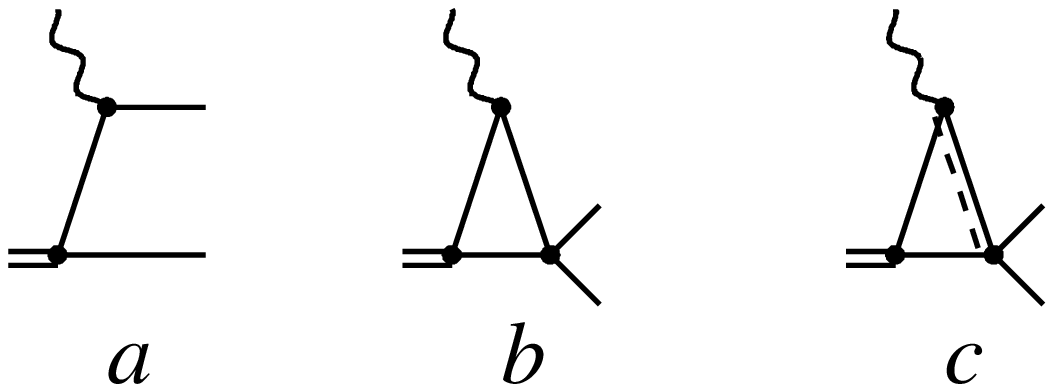}}
\end{center}
\centerline{Fig.1.}
\end{figure}

It should be pointed out that all components of the
photodisintegration amplitude have been calculated in the dispersion
integration technique: 1)~deuteron vertex function which is the
relativistic analogue of the deuteron wave function$^1$; 2)~FSI; its
construction is based on the NN scattering amplitude; 3)~the
amplitudes of diagrams shown in Fig.1.

The main item  of our study is the $NN$ scattering amplitude
$A(s,t)$. This amplitude has been used to find the deuteron vertex
function and FSI$^{1,2}$.

Our dispersion relation approach is based on the dispersion $N/D$
method for $A(s,t)$. In this method the partial wave amplitudes are
treated. Partial amplitudes, $A_l(s)$, in the $s$-channel depend
on $s$ only,  they have all $s$-channel singularities
of $A(s,t)$ (right hand singularities) and left hand singularities,
related to the $t$- and $u$-channel singularities.

The dispersion $N/D$ method provides us an opportunity to
construct  relativistic two-particle partial amplitude in the
region of low and intermediate energies. One can successively
include one open channel after another.

$N_l$-functions are presented as a sum of separable terms
\begin{equation}
N_l(s)=G_l(s)\cdot G_l'(s) + \ldots  ,
\end{equation}
and in our case $G_l'$ differs from $G_l$ in a sign. In the
simplest case the equation for the partial scattering amplitude can
be written in the form$^{1,2}$ \begin{equation}
A_l=G_l^T(1-B_l)^{-1} G_l.  \end{equation}

Here $B_l$ is the dispersion integral for the one-loop diagram:
\begin{equation}
B_l=\int_{4m^2}^{\infty}\frac{ds'}{\pi}\frac{\rho_l') G_l^2(s')}
{s'-s}\,.
\end{equation}
Eq.2 is an analogue of Bethe--Salperter equation for  separable
interaction of a special form.

$G_l$-functions are determined by their left singularities and can be
represented as the integrals along the left hand cut
\begin{equation}
G_l(s)=\int_{-\infty}^{s_L} \frac{ds'}{\pi} \frac{disc G_l(s')}{s'-s}=
\sum_{i=1}^{N_l} \frac{\gamma_{li}}{s-s_{li}}\,.
\end{equation}
To simplify numerical calculations the integral can be replaced
by a sum. The parameters $\gamma_{li},s_{li}$ have been found fitting
the data of the phase shift analysis. The obtained $G$-functions are
used in FSI amplitudes.

In calculations performed in this paper the phase shift analysis of Arndt et
al.$^3$ has been used.

First, the one-channel partial  amplitudes, $A_l(s)$, have been
calculated.  For this case  $G$-functions have been constructed
fitting NN phase shifts, $\delta_{NN}$,  only,  and
photodisintegration amplitude is defined by two diagrams 1a and 1b.

The phase shift analysis by Arndt et al.$^3$ reveals the presence
of an essential inelasticity in the waves $^1S_0$, $^3P_0$,
$^3P_1$, $^3P_2$, $^1D_2$, $^3F_3$ at $T<1$~GeV ( kinetic energy of
the incident proton) .

In the waves $^3P_2$, $^1D_2$, $^3F_3$ the amplitudes have the
resonance-type behaviour. This resonance corresponds to the
intermediate state $N\Delta$. For these waves the two-channel
scattering amplitudes have been built up, and the parameters of
$G$-functions have been found by fitting $NN$ and $N\Delta$ phase
shifts and the transition amplitudes $|A_{NN-N\Delta}|^2$. So, there
is two-channel FSI and photodisintegration amplitude has been
calculated using the diagrams 1a, 1b, 1c. In Fig.3b the partial cross
section ( defined as shown in Fig.2) for the wave $^1D_2$ is
demonstrated. By  definition this cross section is the
contribution of  final state interaction to the total cross
section for different waves. The main contribution is given by the
diagram 1c.

\newpage
\begin{figure}[h]
\begin{center}
\mbox{\epsfxsize=12cm \epsfbox{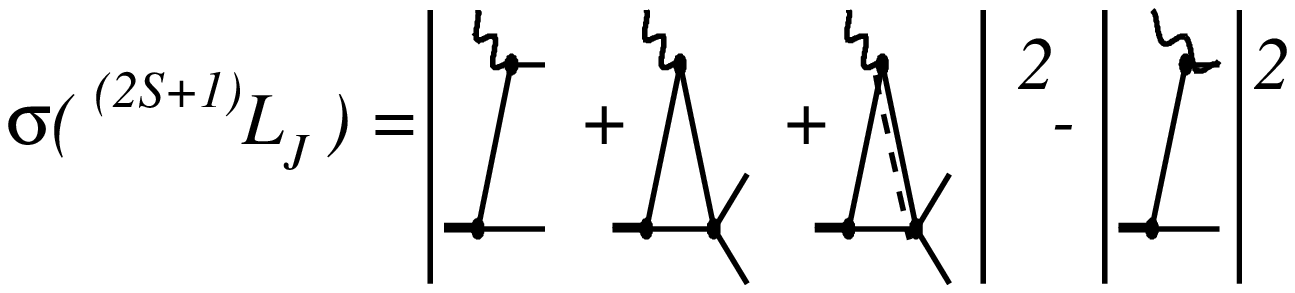}}
\end{center}
\centerline{Fig.2.Definition of the partial cross section.}
\end{figure}

\begin{figure}[h]
\begin{center}
\mbox{\epsfxsize=12cm \epsfbox{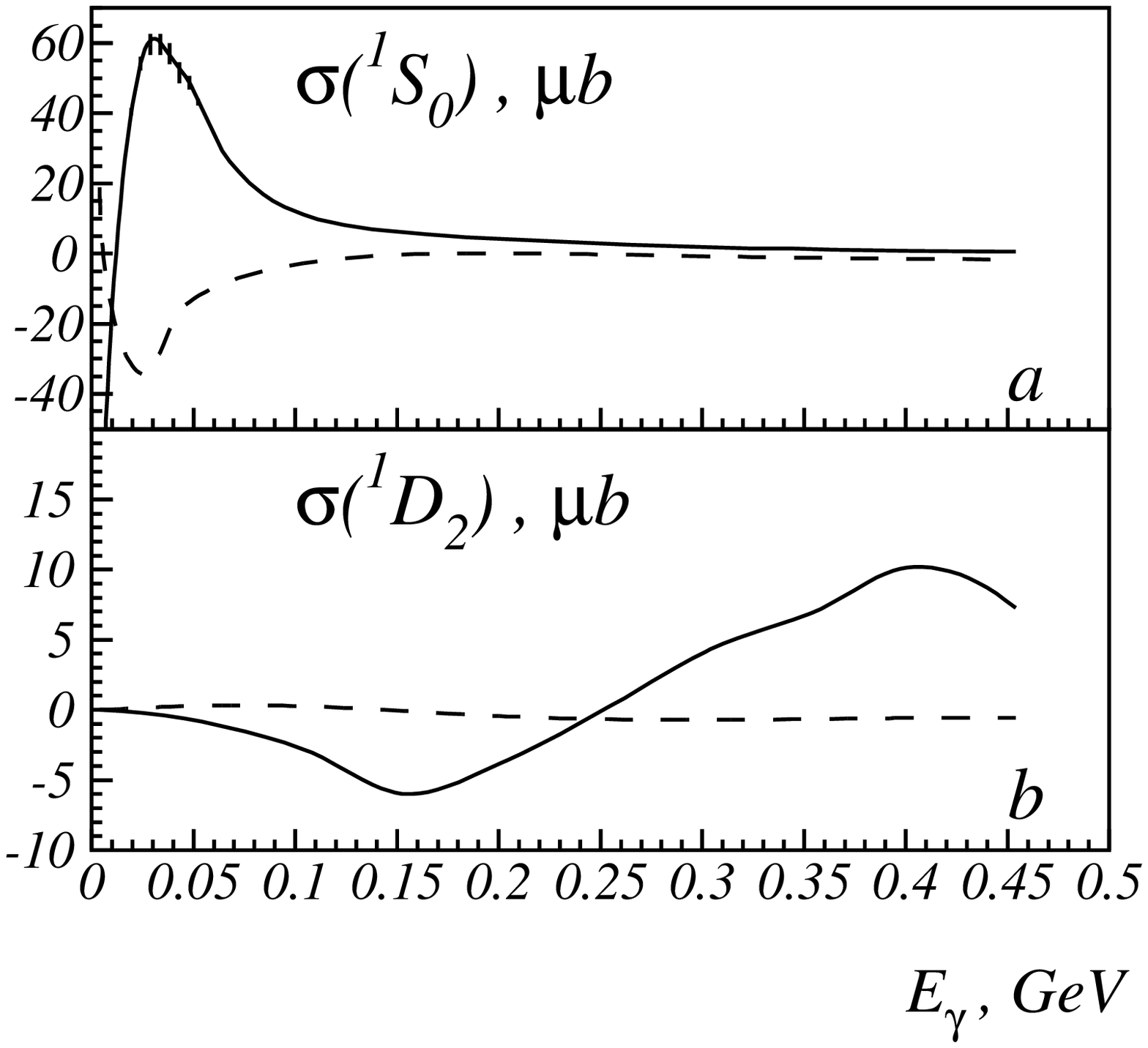}}
\end{center}
Fig. 3. Partial cross sections calculated with (solid line)
 and without (dashed) inelasticity: a) $^1S_0$, b) $^1D_2$.
\end{figure}

For  $^1S_0$, $^3P_0$, $^3P_1$ waves the phase shift analysis
gives  large inelasticity as well, but without resonances in the
amplitudes.  Likewise, the  two-channel amplitude has been
treated, but for the second channel some variants have been tried:
$N\Delta$, $NN^*(1440)$, $\pi N$-pair in the relative $S$-wave. To
find the parameters, phase shifts $\delta_{NN}$ and the parameter of
inelasticity $\rho$ from the paper$^3$ were fitted. In Fig.3a
the partial cross section for the wave $^1S_0$ is demonstrated. The
main contribution  is due to the rescattering processes in the
diagram 1b. The shaded area stands for the different types of
inelasticity.

In Fig.4a  partial cross sections for the waves $^1S_0$, $^3P_0$,
 $^3P_1$ are represented, the largest values of the cross sections
 are for the photon energy $E_\gamma <$100~MeV.

\begin{figure}[h]
\begin{center}
\mbox{\epsfxsize=12cm \epsfbox{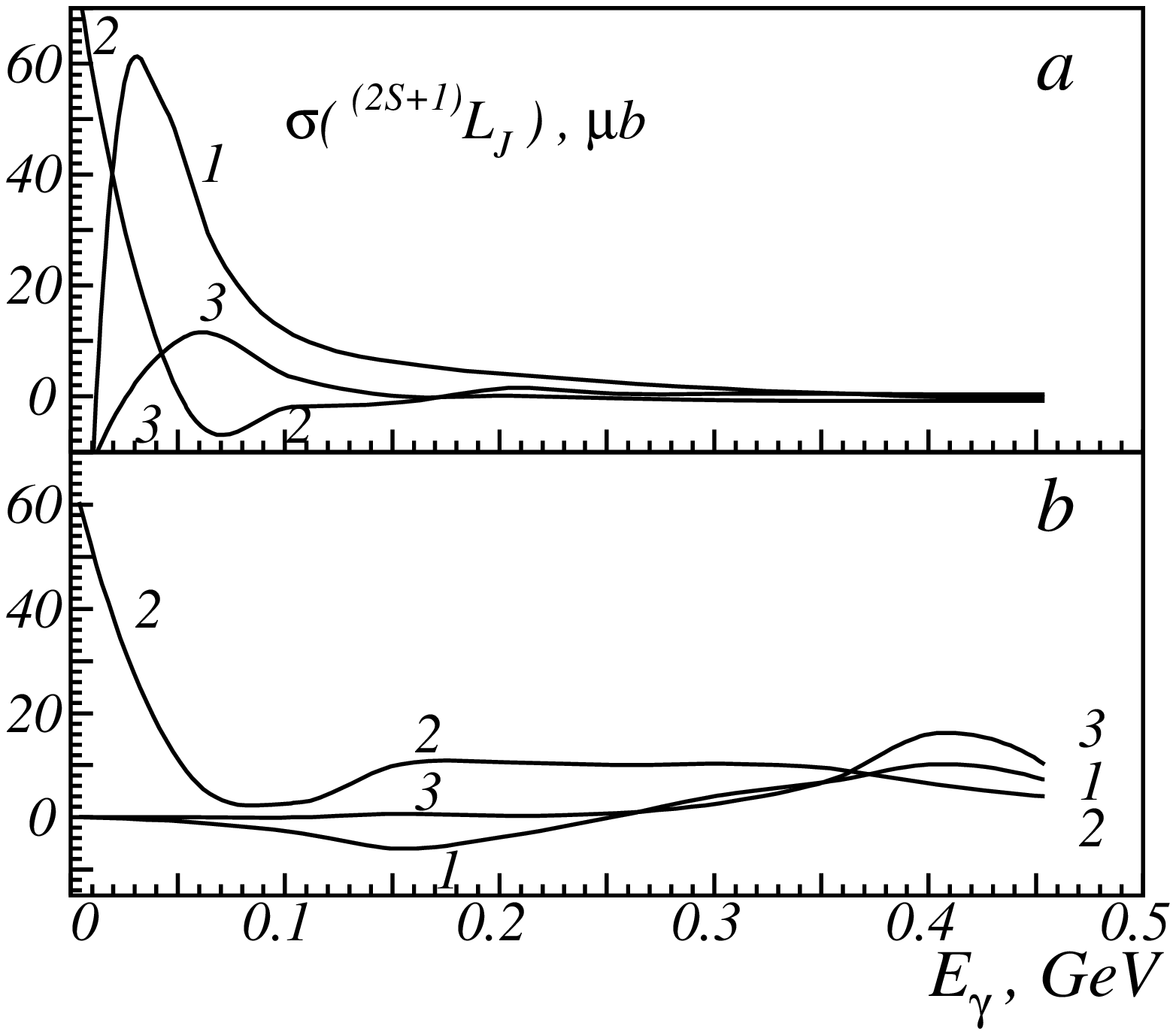}}
\end{center}
Fig. 4. Partial cross sections for the following waves:
a)1: $^1S_0$, 2: $^3P_0$, 3: $^3P_1$; b) 1: $^1D_2$, 2: $^3P_2$,
 3: $^3F_3$.
\end{figure}

 In Fig.4b   partial cross sections for the waves
$^3P_2$, $^1D_2$, $^3F_3$ are demonstrated, which give the main
contribution in the resonance region. We see that the largest
contribution have been obtained from the wave $^3F_3$. In the
well-known nonrelativistic approach of Leidemann and Arenhovel$^4$ the
main contribution in this region is due to the interaction with
the dipole magnetic external field. $M1$ transition from nucleon to
isobar is well known.  This        corresponds mainly to the $^1D_2$
 nucleon-nucleon final state in the diagram 1c.

To understand this, let us turn to the vertex of the isobar
photoproduction. In our calculations the operator introduced
 by Gourdin and Salin$^5$ has been used
\begin{equation}
\Gamma_{\mu\kappa}=c\gamma_5 (\hat q g_{\mu\kappa} -
q_\kappa\gamma_\mu)\,,
\end{equation}
q is  the photon momentum. In the nonrelativistic limit this
 expression has a form
\begin{equation} V(\sigma
',\sigma)=<3/2\sigma'|i(\vec S^+[\vec q,\vec e] )
\left(1-\frac\delta{4m}\right) - \frac\delta{8m} (\vec e T \vec q) +
P(p_\Delta))|1/2 \sigma>\,,
\end{equation}
$\delta=m_\Delta-m$, $\vec e$ is the photon polarization, $\vec
S^+$ is the transition spin and $T$ is the quadrupole transition
operator.

The first term corresponds to the $M1$ transition, the second  one to
$E2$, and there is a part, $P(p_\Delta)$, which depends on the isobar
momentum $(p_\Delta)$.  Due to $P(p_\Delta)$  the photodisintegration
amplitude for the wave $^3F_3$ does not equal to zero.

The operator (5) can be considered as a particular case of the
expression  given by Bjorken and Walecka$^6$. One can tried to check
the results using another operator form for the photodisintegration
vertex.  One of the simplest operators is
\begin{equation}
\Gamma_{\mu\kappa}=ci\varepsilon_{\kappa\mu ab}q_a p_{\Delta b}\,.
\end{equation}
Its nonrelativistic limit is the following

\begin{equation}
V(\sigma ',\sigma)=<3/2\sigma'|i(\vec S^+[\vec q,\vec e])+
\tilde P(p_\Delta)|1/2 \sigma>\,,
\end{equation}
Coupling constants, $c$, for both cases are calculated with the help
of the isobar width$^2$. In Fig.5 one can see that the results for
 partial cross sections differ considerably in these two cases.
Calculations  with   vertex (7) are similar to the
nonrelativistic case.

\begin{figure}[h]
\begin{center}
\mbox{\epsfxsize=12cm \epsfbox{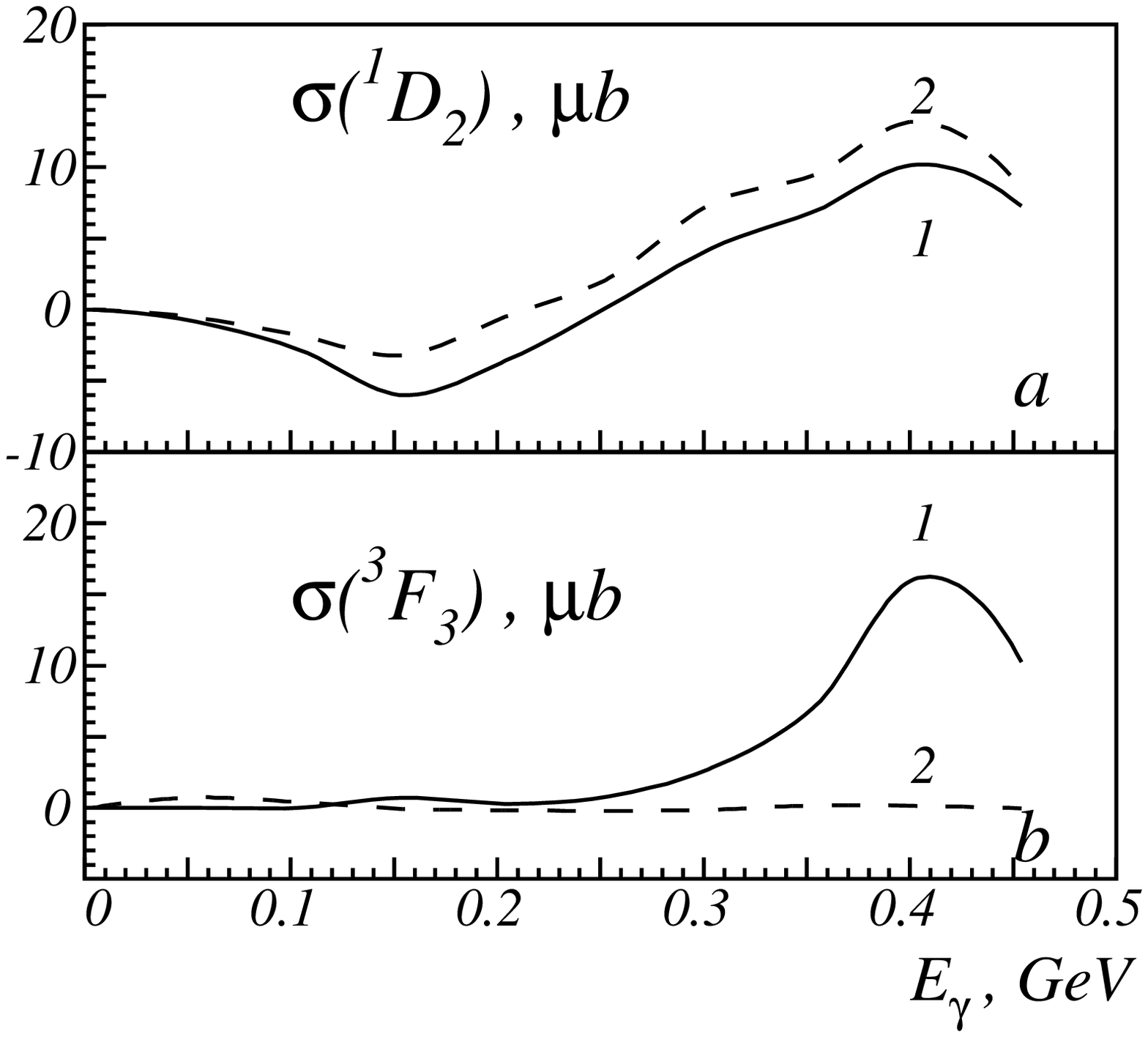}}
\end{center}
Fig. 5. Partial cross sections calculated with vertex operator (5)
 (solid) and (7) (dashed):  a) $^1D_2$; b) $^3F_3$.
\end{figure}

Thus the calculation of  partial cross sections with the correct FSI
is demonstrated.  Special attention should
be paid to the appropriate choice of  the isobar photoproducton
vertex and its form factors.

\section{Acknowledgements}

The author is grateful to A.V.Anisovich and V.V.Anisovich
for helpful discussion.

\end {document}